\documentstyle[aps,prl,epsfig]{revtex}
\begin{document}

\title{Finite Temperature Effective Action in Monopole Background}
\author{Gerald Dunne}
\address{\it Department of Physics, Technion - Israel Institute of
Technology,  Haifa 32000, Israel}
\author{Joshua Feinberg}
\address{Department of Physics, University of Haifa at Oranim, Tivon
36006, Israel \footnote{permanent address}\\
and\\
Department of Physics, Technion - Israel Institute of
Technology,  Haifa 32000, Israel}
\maketitle

\begin{abstract}

We compute the CP-odd part of the finite temperature effective action
for massive Dirac fermions in the presence of a Dirac monopole. We
confirm that the induced charge is temperature dependent, and in the
effective action we find an infinite series of CP-violating terms that
generalize the familiar zero temperature $F\tilde{F}$ term. These results
are analogous to recent results concerning finite temperature induced
Chern-Simons terms. 
\end{abstract}

\vskip 1cm


Recent finite temperature studies \cite{DLL,DGS,FRS,AF,pt,gf,salcedo,hott}
of $2+1$ dimensional Chern-Simons systems \cite{DJT}
have revealed several interesting new features of induced
parity violating terms at finite temperature. It is well known
\cite{redlich,roman} that at T=0, for fermions in the presence of a
static magnetic flux
$\Phi=\int d^2x \frac{eB}{2\pi}$ there is an induced charge
\begin{equation}
Q=-\frac{e}{2} \Phi
\label{cscharge}
\end{equation}
and this corresponds to an induced Chern-Simons term in the zero
temperature Euclidean effective action:
\begin{equation}
S_{P-odd}=-i \frac{e^2}{8\pi}\int d^3 x\, \epsilon^{\mu\nu\rho}A_\mu
\partial_\nu A_\rho=-i \frac{\Phi}{2}\,\int e A_0\, d\tau
\label{cs0}
\end{equation}
But at finite temperature, in such a static background, the induced
charge is \cite{BDP}
\begin{equation}
Q=-\frac{e}{2}\Phi\, \tanh(\frac{\beta m}{2})
\label{cscharget}
\end{equation}
where $\beta=\frac{1}{T}$ is the inverse temperature. The corresponding
parity-odd part of the (Euclidean) finite temperature induced effective action
is \cite{DLL,DGS,FRS,AF,pt,gf,salcedo,hott}
\begin{eqnarray}
S_{P-odd}&=& -i \Phi\, {\rm arctan}\left[\tanh(\frac{\beta m}{2})\, 
\tan(\frac{1}{2}\int_0^\beta e A_0 d\tau)\right]\nonumber\\
&=& -i \Phi\left[\tanh(\frac{\beta m}{2})(\frac{1}{2}\int_0^\beta e A_0
d\tau)+\frac{1}{3}\tanh(\frac{\beta m}{2}){\rm sech}^2(\frac{\beta m}{2})
(\frac{1}{2}\int_0^\beta e A_0 d\tau)^3 +\dots \right]
\label{cst}
\end{eqnarray}
The perturbative expansion in the second line of (\ref{cst}) shows
that at nonzero temperature there is an infinite series of parity
violating terms, of which the Chern-Simons term, with temperature
dependent charge (\ref{cscharget}), is only the first. The entire series
is required at finite $T$ to show that
$S_{P-odd}$ is invariant under ``large" gauge transformations $A_0\to
A_0+\frac{2\pi}{e\beta} N$, $N\in {\bf Z}$
\cite{DLL,DGS,FRS,AF,pt,gf,salcedo,hott}. The higher terms are
non-extensive (i.e. they are not integrals of a Lagrangian density), but
they all vanish at zero temperature, leaving just the Chern-Simons term
(\ref{cs0}). 

In this Letter we investigate similar phenomena for the finite
temperature induced charge and induced effective action in $3+1$
dimensional field  theory. The analogue
discrete symmetry we consider is CP, and we compute the finite $T$ induced
charge, and the corresponding finite $T$ induced effective action, for
massive Dirac fermions in the presence of a Dirac monopole. This should be
distinguished from earlier work \cite{rw}, in a similar spirit, concerning
CP-violating terms at high temperature and finite fermion density in the
effective action for an even number of (massless) chiral fermions.

We recall that the
quantum mechanics of a  particle of electric charge $e$ and a magnetic
pole of magnetic charge $g=\frac{1}{4\pi}\int d^3x
\vec{\nabla}\cdot\vec{B}$, is consistent only if these charges satisfy
\cite{dirac,wy,kazama}
\begin{equation}
2 e g\equiv 2q  ={\rm integer}
\label{dirac}
\end{equation}
But the Dirac Hamiltonian 
\begin{equation}
H_D=-\vec{\alpha}\cdot(i\vec{\nabla}+e \vec{A}_D)+m\gamma_0
\label{dh}
\end{equation}
for a fermion in the presence of a Dirac monopole $\vec{A}_D$ is
not self-adjoint, and so requires the introduction of a self-adjoint
extension parameter $\theta$ that characterizes the boundary conditions
(in the s-wave sector) at the location of the monopole
\cite{goldhaber,callias,yamagishi,dhoker}.
The spectrum of the fermions
becomes
$\theta$ dependent, and there is an induced charge of the fermion vacuum
given by \cite{W,wilczek,grossman}
\begin{equation}
Q=-\frac{e\theta}{2\pi} 2q
\label{witten}
\end{equation}
This induced charge corresponds to an induced CP-violating term in the
zero temperature Euclidean effective action \cite{W,wilczek}:
\begin{equation}
S_{CP-odd}=i \frac{\theta e^2}{16 \pi^2}\int d^4 x \,
F_{\mu\nu}\tilde{F}_{\mu\nu}=-i \frac{e^2\theta}{4\pi^2} \int d^4 x\,
A_0\,\vec{\nabla}\cdot \vec{B} =-i 2q\frac{\theta}{2\pi} \int e A_0d\tau
\label{mon0}
\end{equation}
The finite temperature induced charge for fermions in the presence of a
Dirac monopole has been computed recently \cite{cp,gps}, and the zero
temperature result (\ref{witten}) becomes temperature dependent: 
\begin{equation}
Q=- e \frac{x}{\pi} (2q) \sin\theta\, \sum_{n=0}^\infty {1\over
(2n+1)^2+x^2+x \cos\theta\sqrt{(2n+1)^2+x^2}}
\label{coriano}
\end{equation}
where $x\equiv \frac{m\beta}{\pi}$. In \cite{cp} it was checked that the
zero temperature limit (i.e.
$x\to\infty$) of the expression (\ref{coriano}) agrees with the zero
temperature result (\ref{witten}) [see also (\ref{q0}) below].

These results should be compared with the related work of Le Guillou and
Schaposnik \cite{ls}, who considered the problem of the electric charge of a
non-Abelian dyon at finite T with CP violation introduced by an explicit
$F \tilde{F}$ term, but no fermions. There, Witten's relation
(7) was found to be unchanged at finite T. However, as discussed in \cite{gps}, 
the analysis of \cite{ls} was effectively at tree level, and the inclusion of light
fermions was shown \cite{gps} (in the limit of an infinitely
massive 't Hooft-Polyakov monopole) to make the charge temperature
dependent in a manner analogous to (9).

In this Letter we confirm the result (\ref{coriano}) using a
slightly different approach, and we then compute the corresponding CP-odd
part of the finite temperature induced effective action for fermions
in the background of a static Dirac monopole. For this purpose it is
convenient to express everything in terms of the spectral function (the
density of states) of the Dirac equation for fermions in the background
of a static Dirac monopole. This
spectral function is computed in the appendix using an elementary
calculation of the resolvent of the (s-wave) Dirac Hamiltonian. This leads
to a simple integral representation of the summation formula
for the charge in (\ref{coriano}). We then obtain a
spectral representation of the CP-odd part of the induced effective
action at finite temperature, and study its dependence on temperature
and on $\theta$. 


The finite temperature induced charge is expressed in terms of
the spectral function as \cite{niemi}
\begin{equation}
Q=-\frac{e}{2} \int_{-\infty}^\infty d\omega\, \rho_D(\omega)\,
\tanh(\frac{\beta\omega}{2})
\label{charget}
\end{equation}
The spectral function $\rho_D(\omega)$ is defined in the usual manner as 
\begin{equation}
\rho_D(\omega)=\frac{1}{\pi} {\cal I}m \, {\rm Tr} \left({1\over
H_D-\omega-i\epsilon}\right) \equiv \frac{1}{\pi}\, {\cal I}m\,
\Gamma_D(\omega+i\epsilon)
\label{spectral}
\end{equation}
where $H_D$ is the Dirac Hamiltonian in (\ref{dh}), and (\ref{spectral})
defines $\Gamma_D(\omega)$, the trace of the resolvent of $H_D$.

In all but the lowest partial wave, the spectrum of the Dirac
Hamiltonian $H_D$ has a symmetry between positive and negative energy
states \cite{goldhaber,yamagishi,dhoker}, so that only the lowest partial
wave contributes to the induced charge in (\ref{charget}). Thus we only
need to consider this lowest partial wave sector, which is effectively a
one dimensional radial problem (i.e. one dimensional, but on the half-line
$r\geq 0$). The effective one dimensional Hamiltonian is 
\cite{goldhaber,callias,yamagishi,dhoker} (we choose $q>0$)
\begin{equation}
H=-i\gamma_5 \,\frac{d}{dr}+m\,\gamma_0 
\label{ham}
\end{equation}
acting on two-component spinors $\chi(r)$ with the ``bag" boundary
condition
\begin{equation}
i \gamma_1 \chi(0)=\exp(-i\gamma_5 \theta)\chi(0)
\label{bag}
\end{equation}
Here, the $2\times 2$ Dirac matrices
are:
\begin{equation}
\gamma_0=\left(\matrix{1&0\cr 0&-1}\right)\qquad
\gamma_1=\left(\matrix{0&1\cr -1&0}\right) \qquad
\gamma_5=\left(\matrix{0&1\cr 1&0}\right)
\label{mat}
\end{equation}
The boundary condition (\ref{bag}) ensures self-adjointness of the
reduced Hamiltonian $H$ in (\ref{ham}), with the spinor inner product
being $<\chi|\psi>=\int_0^\infty dr\, \chi^\dagger (r) \psi(r)$. We also
note here that we can restrict our attention to $0<\theta <\pi$, since a
CP-inversion takes $\theta\to -\theta$, and CP-odd observables are
correspondingly odd in $\theta$.

Given that only this lowest partial wave contributes to the induced
charge, we obtain from (\ref{charget}) and (\ref{spectral}) the following
contour integral representation of the induced charge
\begin{equation}
Q=-\frac{e}{2}\, (2q)\, \oint_C \frac{dz}{2\pi i}\, \tanh(\frac{\beta
z}{2})\, \Gamma(z)
\label{contour}
\end{equation}
where C is the contour in Figure 1, and $\Gamma(z)$ is the trace of the
resolvent of the reduced Hamiltonian $H$. The overall factor of
$2q$ arises as the degeneracy of the lowest partial wave ({\it viz},
$2j+1$, with $j=q-\frac{1}{2}$).

In the appendix, we present an elementary derivation of $\Gamma(z)$ in
terms of the diagonal resolvent (i.e. diagonal Green's function)
$\Gamma(r;z)$ of $H$ [see Eqs. (\ref{diag},\ref{final})]:
\begin{equation}
\Gamma(z)\equiv\int_0^\infty dr \,\Gamma(r;z) = -\frac{m}{2k^2}
\left({z\,\sin\theta-m-ik\,\cos\theta\over z-m\,\sin\theta}\right)
\label{spec}
\end{equation}
Here $z^2=k^2+m^2$, and ${\cal I}m\,k>0$ on the physical sheet, and we have 
dropped a trivial $\theta$-independent term from $\Gamma(z)$ (see the 
appendix). Notice that
for (real) infinitesimal $\lambda$ and $\epsilon$,
\begin{equation}
\Gamma(m\, \sin\theta+\lambda+i\epsilon)=-\frac{\Theta(-\cos\theta)}{\lambda
+i\epsilon}
\label{pole}
\end{equation}
and thus, in addition to scattering states, there is a
bound state at $z=m\,\sin\theta$, provided
$\cos\theta<0$. Using the
result (\ref{spec}) and deforming the contours around the cuts and around the 
bound
state pole (which is only present if $\cos\theta<0$) we obtain the
following expression for the induced charge as a real integral:
\begin{equation}
Q=-\frac{qe}{\pi} \left[\pi\,\tanh(\frac{m\beta}{2}\,\sin\theta)
\Theta(-\cos\theta)+\sin\theta\, \cos\theta\, \int_1^\infty {du\over
\sqrt{u^2-1}} \left({\tanh(\frac{m\beta}{2}\,u)\over
u^2-\sin^2\theta}\right)\right]
\label{qint}
\end{equation}
We plot this induced charge in Figure 2 as a function of $\theta$ for
various values of $m\beta$, and in Figure 3 as a function of $m\beta$ for
various values of $\theta$. It is clear from these Figures that in the
zero temperature limit ($m\beta\to\infty$), $-\frac{\pi Q}{eq}$ saturates
to $\theta$, in agreement with the zero temperature result (\ref{witten}).
This is easy to verify analytically from the integral representation
(\ref{qint}), since
\begin{eqnarray}
Q(T=0)&=&-\frac{q e}{\pi} \left[\pi\,\Theta(-\cos\theta) +
\sin\theta\,\cos\theta\, \int_1^\infty {du\over
\sqrt{u^2-1}} \left({1\over u^2-\sin^2\theta}\right)\right]\nonumber\\
&=&-\frac{q e}{\pi} \, \theta
\label{q0}
\end{eqnarray}
In the high temperature limit ($m\beta\to 0$), the induced charge
vanishes for any $\theta$, as is clear from Figures 2 and 3. 
Analytically,
\begin{eqnarray}
Q(T\to\infty)&\sim &-\frac{q e}{\pi}  \left[\pi
\frac{m\beta}{2}\,\sin\theta\,\Theta(-\cos\theta) +
\frac{m\beta}{2} \sin\theta\,\cos\theta\, \int_1^\infty {du\over
\sqrt{u^2-1}} \left({u\over u^2-\sin^2\theta}\right)\right]\nonumber\\
&=&- \frac{q e m}{4T} \, \sin\theta
\label{qinf}
\end{eqnarray}
When $\theta=\frac{\pi}{2}$, the finite temperature induced charge 
(\ref{qint}) reduces to the simple form:
\begin{equation}
Q(\theta=\frac{\pi}{2})=-\frac{q e}{2} \, \tanh(\frac{m\beta}{2})
\label{qhalf}
\end{equation}
As mentioned in \cite{cp}, this is consistent with the $2+1$ dimensional
result (\ref{cscharget}), since it is known \cite{gerbert} that for
$\theta=\frac{\pi}{2}$, the monopole problem with $2q=1$ reduces to that
of an Aharonov-Bohm flux string with flux $\Phi=\frac{1}{2}$. 

These results (\ref{qint},\ref{q0},\ref{qinf},\ref{qhalf}) are in complete
agreement with the summation expression (\ref{coriano}) for the finite
temperature induced charge found by Coriano and Parwani in
\cite{cp}. Indeed, the integral expression (\ref{qint}) is in fact the
Sommerfeld-Watson representation of the sum in (\ref{coriano}); and
equivalently, the sum in (\ref{coriano}) can be obtained from
(\ref{contour}) by deforming the contour around the poles of the
$\tanh(\frac{\beta z}{2})$ function on the imaginary $z$ axis. The main
utility of the integral representation (\ref{qint}), as compared to the
summation representation (\ref{coriano}), becomes apparent when we compute
the CP-odd part of the finite temperature induced effective action.


To compute the finite temperature effective action, we work in Euclidean
space and consider fermions in the presence of the
background gauge field $A_\mu=(A_0,\vec{A}_D)$, where $A_0$ is a constant
(this can always be achieved by a small gauge transformation), and
$\vec{A}_D$ is the static vector potential for a Dirac monopole. The
finite temperature induced effective action is
\begin{equation}
S=\int_{-\infty}^\infty d\omega \, \rho_D(\omega)\, {\rm log}\, {\rm
cosh}\left(\frac{\beta}{2}(\omega-ie A_0)\right)
\label{fulleff}
\end{equation}
As we are only interested in the CP-odd part of the
induced effective action, two simplifications occur. First, since
only the lowest partial wave sector of the spectrum has a 
CP-odd piece, we can (just as we did for the charge) use the reduced
spectral function $\rho(\omega)$ of the s-wave Hamiltonian (\ref{ham}),
rather than the full spectral function $\rho_D(\omega)$ of the Dirac
Hamiltonian $H_D$ in (\ref{dh}). Second, since the CP-odd part must be
odd in $A_0$, we can compute the difference as:
\begin{eqnarray}
S_{CP-odd}&=&\frac{1}{2}\, (2q)\int_{-\infty}^\infty d\omega \,
\rho(\omega)\, {\rm log}\left[{ {\rm cosh}\left(\frac{\beta}{2}(\omega-ie
A_0)\right)\over {\rm cosh}\left(\frac{\beta}{2}(\omega+ie
A_0)\right)}\right]\nonumber\\
&=&-iq\int_{-\infty}^\infty d\omega \, \rho(\omega)\,
\left[\tanh(\frac{\beta \omega}{2})(e\beta A_0)+\frac{1}{12}
\tanh(\frac{\beta \omega}{2}){\rm sech}^2(\frac{\beta\omega}{2}) (e\beta
A_0)^3+\dots \right]\nonumber\\
\label{cpodd}
\end{eqnarray}
In this expansion, the first term, which is linear in $A_0$,
corresponds to the familiar zero temperature induced effective
action in (\ref{mon0}), but with a temperature dependent induced 
charge (\ref{contour}) multiplying $A_0$. The higher order terms in the
perturbative expansion (\ref{cpodd}) of the CP-odd part of the induced
effective action are not of this $\int F\tilde{F}$ form. They are
non-extensive in Euclidean time, as they involve powers of $e\beta
A_0=(e\int_0^\beta A_0\,d\tau)$. Nevertheless, all these non-extensive
terms vanish at zero temperature due to factors of ${\rm
sech}^2(\frac{\beta \omega}{2})$, which vanishes exponentially fast as
$T\to 0$. Thus, at zero temperature only the first term survives, and the
CP-odd part of the finite temperature induced effective action
(\ref{cpodd}) reduces to the familiar zero temperature expression
(\ref{mon0}). 

Given that we know the relevant spectral function $\rho(\omega)$, it is
straightforward to write down an integral representation of any term in
the perturbative expansion in (\ref{cpodd}). The coefficient of
$(i\int_0^\beta A_0 \,d\tau)$ is just the induced charge (\ref{qint}),
while the coefficient of $i(\int_0^\beta A_0 \,d\tau)^3$ is
\begin{equation}
J\equiv -\frac{e^3 q}{12\pi} \left[\pi\tanh(\frac{m\beta\sin\theta}{2})
{\rm sech}^2(\frac{m\beta\sin\theta}{2})
\Theta(-\cos\theta)+\sin\theta\, \cos\theta\, \int_1^\infty {du\over
\sqrt{u^2-1}} {\tanh(\frac{m\beta u}{2}) {\rm sech}^2
(\frac{m\beta u}{2})
\over u^2-\sin^2\theta}\right]
\label{first}
\end{equation}
In Figure 4 we plot this first correction term $J$ as a
function of
$m\beta$, for various values of $\theta$. It is clear that $J$ vanishes
exponentially fast as
$m\beta\to\infty$ (i.e. as $T\to 0$). Using the integral representation
(\ref{first}), we find that (for $\theta\neq \frac{\pi}{2}$)
\begin{equation}
J(T\to 0,\theta\neq\frac{\pi}{2} )\sim -\frac{e^3 q}{12\pi}\left[4\pi
e^{-m\beta\sin\theta}\,\Theta(-\cos\theta) +2 \sqrt{2\pi}
\tan\theta\, {e^{-m\beta}\over \sqrt{m\beta}}\right]\quad, \qquad
m\beta\to\infty
\label{asymptotic}
\end{equation}
When $\theta=\frac{\pi}{2}$ we can evaluate $J$ exactly:
\begin{eqnarray}
J(\theta=\frac{\pi}{2} )&=& -\frac{e^3 q}
{24}\,\tanh(\frac{m\beta}{2}) {\rm
sech}^2(\frac{m\beta}{2})\nonumber\\ &\sim& -\frac{e^3 q}{6}\,
e^{-m\beta}\quad, \qquad m\beta\to\infty
\label{special}
\end{eqnarray}
These asymptotic formulae (\ref{asymptotic},\ref{special}) match the large
$m\beta$ behaviour in Figure 4, but note that when $\theta\neq
\frac{\pi}{2}$, the  sign of
$\cos\theta$ determines which exponential factor dominates in
(\ref{asymptotic}). 

At high temperatures ($m\beta\to 0$), 
\begin{eqnarray}
J(T\to\infty)&\sim& -\frac{e^3 q}{12\pi}
\left[\pi\frac{m\beta\sin\theta}{2}
\Theta(-\cos\theta)+\frac{m\beta}{2} \sin\theta\, \cos\theta\,
\int_1^\infty {du\over
\sqrt{u^2-1}} {u\over u^2-\sin^2\theta}\right]\nonumber\\
&=& -\frac{e^3 q m}{48 T}\, \sin\theta
\label{jhight}
\end{eqnarray}
This involves the same $\frac{\sin\theta}{T}$ factor as the high
temperature limit of the induced charge $Q$ in (\ref{qinf}). In fact,
this is true of the high $T$ behaviour of every term in the expansion in
(\ref{cpodd}), so we find that as $T\to\infty$, the CP-odd
induced effective action behaves as
\begin{equation}
S_{CP-odd}\sim -i 
\left(\frac{qm}{2T}\right) \,\sin\theta\,\tan(\frac{e}{2}\int_0^\beta A_0)
\label{hight}
\end{equation}
Finally, we comment on some special values of $\theta$. When $\theta=0$
or $\pi$, the CP-odd finite temperature effective action (\ref{cpodd})
vanishes, as is easy to verify term by term in the perturbative
expansion. This is of course consistent with the fact that CP is not
violated for these values of $\theta$ \cite{goldhaber,callias,yamagishi,dhoker}.
When $\theta=\frac{\pi}{2}$, we can evaluate every term in the
perturbative expansion in (\ref{cpodd}). Each term contributes one half
of the bound state contribution, so we find the simple closed form
expression:
\begin{equation}
S_{CP-odd}(\theta=\frac{\pi}{2})=-i\,q\, {\rm arctan}\left[\tanh(\frac{\beta m}{2})\, 
\tan(\frac{1}{2}\int_0^\beta e A_0 d\tau)\right]
\label{pi2}
\end{equation}
Note that this is {\it exactly} the same form as the finite temperature
induced parity-odd effective action (\ref{cst}) found in $2+1$
dimensions, with the identification $q=\Phi$. This is consistent with the
fact \cite{gerbert,cp} that the s-wave spectral properties of the 
$\theta=\frac{\pi}{2}$ fermion plus monopole system coincide with those of
the planar fermion plus Aharonov-Bohm string system with flux $\Phi=q$.


To conclude, we note that it would
be of interest to extend this analysis to fermions in the presence of a 't
Hooft-Polyakov monopole. The finite temperature induced charge has
been computed \cite{gps} in the limit of an infinitely massive
monopole coupled to light fermions, but it would be interesting to try to
go beyond this point monopole limit. For the effective action, another
interesting question is to go beyond the {\it static} background ansatz,
to find other structures that generalize the zero temperature
$F\tilde{F}$ term for nonzero temperature. Even in the simpler
Chern-Simons case, the calculation of induced terms at finite temperature
is much more complicated for a non-static background
\cite{kao,DD1}.

\section*{Appendix}

In this appendix we present an elementary derivation of the spectral
function for the reduced s-wave hamiltonian $H$ in (\ref{ham}). The
resolvent (or Green's function) $G(r,r^\prime;z)$ for $H$ is defined
by:
\begin{equation}
(H- z)G(r,r^\prime;z)=\delta(r-r^\prime) {\bf 1}
\label{green}
\end{equation}
This is a $2\times 2$ matrix equation, so we write 
\begin{equation}
G(r,r^\prime)=\left(\matrix{a(r,r^\prime)&b(r,r^\prime)\cr c(r,r^\prime) &
d(r,r^\prime)}\right)
\label{decomp}
\end{equation}
where we have suppressed the $z$ dependence. Using the Hamiltonian in
(\ref{ham}), we find that $a$ and $c$ satisfy:
\begin{eqnarray}
-{\partial_r^2 a(r,r^\prime)\over z+m}-(z-m)\, a(r,r^\prime)&=&\delta
(r-r^\prime)\nonumber\\
c(r,r^\prime)&=&-i\, {\partial_r a(r,r^\prime)\over z+m}
\label{ac}
\end{eqnarray}
with $d$ and $b$ satisfying similar equations as $a$ and $c$
(respectively) with
$m\to -m$. It is trivial to solve these equations in terms of the
fundamental solutions
\begin{equation}
\psi_\pm(r)=e^{\pm i k r}
\label{fund}
\end{equation}
where $z^2=m^2+k^2$, and ${\cal I}m (k)>0$ on the physical sheet. We find
\begin{equation}
a(r,r^\prime)=-\left(\frac{z+m}{2ik}\right) \left[\Theta(r-r^\prime)
e^{ik(r-r^\prime)} +\Theta(r^\prime-r) e^{-ik(r-r^\prime)} + \alpha\, 
e^{ik(r+r^\prime)}\right]
\label{exp}
\end{equation}
where the last term is a solution of the homogeneous equation. (Note that
the homogeneous solution $e^{-ik(r+r^\prime)}$ is excluded by its large
$r$ behaviour for $r=r^\prime$ and ${\cal I}m (k)>0$). The coefficient
$\alpha$ in (\ref{exp}) is fixed by the boundary condition (\ref{bag})
which, when applied to the resolvent $G(r,r^\prime)$, requires that 
\begin{equation}
a(0,r^\prime)=i\left(\frac{1+\sin\theta}{\cos\theta}\right)\,
c(0,r^\prime)\quad ,\quad r^\prime >0
\label{bagres}
\end{equation}
This determines the constant $\alpha$ to be:
\begin{equation}
\alpha={z\, \sin\theta-m-ik\,\cos\theta\over z-m\,\sin\theta}
={k\,\sin\theta-i\,z\,\cos\theta\over k+i\,m\,\cos\theta}
\label{constant}
\end{equation}
Thus, the diagonal part of $a(r,r^\prime)$ is
\begin{equation}
a(r,r)=-\left(\frac{z+m}{2ik}\right)\left(1+\alpha\, e^{2ikr}\right)
\label{adiag}
\end{equation}
A similar calculation for $d(r,r^\prime)$ leads to
\begin{equation}
d(r,r)=-\left(\frac{z-m}{2ik}\right)\left(1-\alpha\, e^{2ikr}\right)
\label{ddiag}
\end{equation}
Thus, the diagonal resolvent is
\begin{equation}
\Gamma(r;z)\equiv {\rm Tr}\left[G(r,r;z)\right]= -\frac{1}{ik}\left[z+
\alpha\, m\, e^{2ikr}\right]
\label{diag}
\end{equation}
The first term in (\ref{diag}) corresponds to the free case and may be
dropped (also it is independent of $\theta$ and so does not contribute
to induced CP-violating terms). Thus, the trace of the diagonal resolvent
is
\begin{equation}
\Gamma(z)\equiv \int_0^\infty dr\, \Gamma(r;z)=-\frac{\alpha m}{2k^2}
\label{final}
\end{equation}
with $\alpha$ given in (\ref{constant}). Note that $\alpha$ has a pole at
$k=-im\cos\theta$ in the $k$ upper half plane provided $\cos\theta<0$.
Thus, when
$\cos\theta<0$ there is a bound state of energy
$z_b=m\,\sin\theta$, in addition to scattering states.

\section*{Acknowledgments}

Both authors acknowledge the support of the Israeli Science Foundation
through grant number 307/98(090-903). GD acknowledges the
support of the U.S. DOE under grant DE-FG02-92ER40716.00, and thanks the
Technion Physics Department for its hospitality.

\begin{figure}[h]
\centering{\epsfig{file=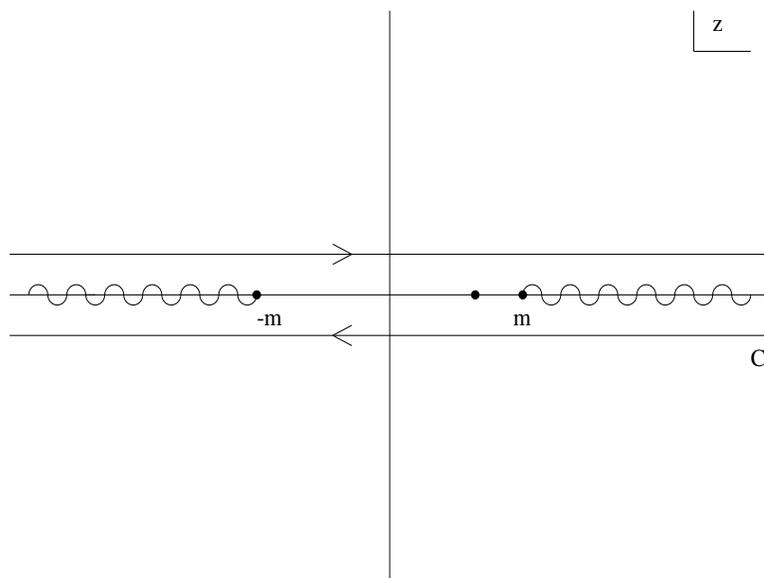,width=4in,height=3in}}
\caption{Contour of integration $C$, in the complex energy plane, for the
induced charge in (\protect{\ref{contour}}). Note the branch cuts
beginning at the continuum thresholds $z=\pm m$, and the bound state pole
at $z=m\,\sin\theta$, which is present only if $\cos\theta <0$.}
\end{figure}

\begin{figure}[h]
\centering{\epsfig{file=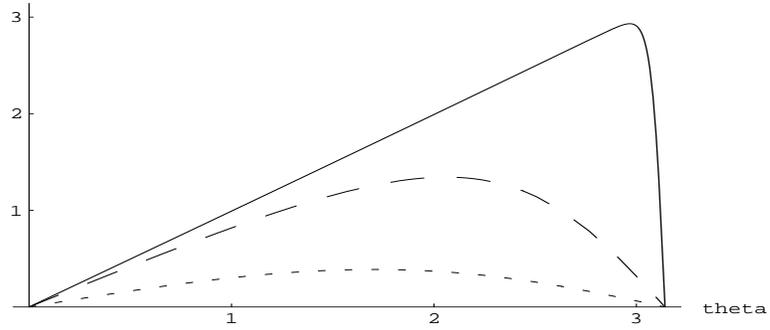,width=4in,height=3.5in}}
\vskip -2cm
\caption{Plot of $-\frac{\pi Q}{qe}$ in (\protect{\ref{qint}}) as a function
of $\theta$ for various values of $m\beta=\frac{m}{T}$. The short-dash
curve is for $m\beta=0.5$, the long-dash curve is for $m\beta=2$, and the
solid curve is for $m\beta=30$. As $T\to 0$, the plot tends to $\theta$,
while at high $T$ it is proportional to $\sin\theta$. This type of plot, based
on the summation formula (\protect{\ref{coriano}}) appeared in the work of
Coriano and Parwani; here for completeness we have 
re-plotted it using the integral representation (\protect{\ref{qint}}) of the 
finite $T$ induced charge.} 
\end{figure}

\begin{figure}[h]
\centering{\epsfig{file=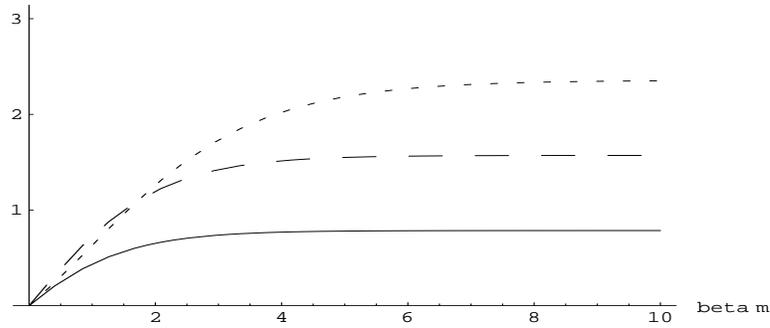,width=4in,height=3.5in}}
\caption{Plot of $-\frac{\pi Q}{qe}$ in (\protect{\ref{qint}}) as a function
of $m\beta=\frac{m}{T}$ for various values of $\theta$. The solid
curve is for $\theta=\frac{\pi}{4}$, the long-dash curve is for
$\theta=\frac{\pi}{2}$, and the short-dash curve is for
$\theta=\frac{3\pi}{4}$. As
$T\to 0$, each curve saturates rapidly to $\theta$.}
\end{figure}

\begin{figure}[h]
\centering{\epsfig{file=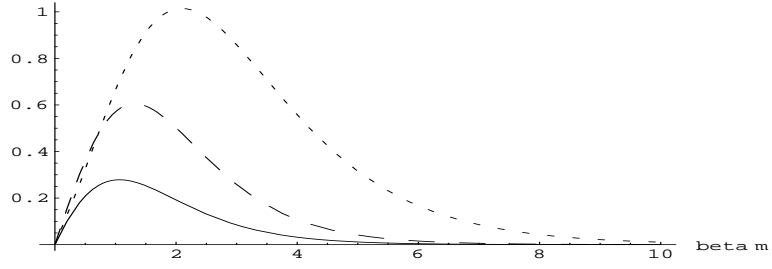,width=4in,height=3in}}
\caption{Plot of $-\frac{12\pi}{qe^3}J$ in (\protect{\ref{first}}) as a
function of $m\beta=\frac{m}{T}$ for various values of $\theta$. The solid
curve is for $\theta=\frac{\pi}{4}$, the long-dash curve is for
$\theta=\frac{\pi}{2}$, and the short-dash curve is for
$\theta=\frac{3\pi}{4}$. Note the
exponential decay at low temperatures.}
\end{figure}

\end{document}